# Hierarchical XP

## Improving XP for large scale projects in analogy to reorganization processes

Carsten Jacobi
Lehrstuhl für Betriebswirtschaftslehre
Munich University of Technology
Germany
http://www.bwl.wiso.tu-muenchen.de

Bernhard Rumpe
Software & Systems Engineering
Munich University of Technology
Germany
http://www.rumpe.de/

**Abstract.** XP is a light-weight methodology suited particularly for small-sized teams that develop software which has only vague or rapidly changing requirements. The discipline of systems engineering knows it as approach of incremental system change or also of "muddling through". In this paper, we introduce three well known methods of reorganizing companies, namely, the holistic approach, the incremental approach, and the hierarchical approach. We show similarities between software engineering methods and company reorganization processes. In this context, we discuss the extreme programming (XP) approach and how the elements of the hierarchical approach can improve XP. We provide hints on how to scale up XP to larger projects e.g. those common in the telecommunication industry.

## 1. INTRODUCTION

Extreme Programming (XP) [Beck99] is a light-weight methodology for small-sized teams developing software with vague or rapidly changing requirements. XP can be regarded as an explicit reaction to the complexity of today's modeling techniques like the Unified Process [Jacobson+99], the V-model [KBSt97], Catalysis [D'Souza+98], or the Open Modeling Language [Firesmith+97]. XP focuses especially on a small number of best practices, disregarding many others advocated by the other methodologies. Being a relatively new, promising approach, XP still has to be proven in various projects on its superiority and usefulness when compared to other approaches. Measuring the quality of a software development in quantitative numbers is such a complex issue, so that we will mainly have a common sense of the advantages of XP over other methods. XP will evolve, reducing its weaknesses and increasing its strengths. This article is not a criticism of XP, but suggests an improvement in one of its obvious weaknesses: XP is designed for a single small team of less than a dozen team members, therefore, it has its problems to scale up for larger projects. Fortunately, applying the XP approach in projects seems to considerably downsize the number of necessary participants, but there is still a number of areas, where hundreds of developers work on producing one single software product. For example, the telecommunication industry is under enormous pressure to add and improve functionality of their products. The time to market span in the mobile phone business needs fast and flexible process for large projects. Switching systems need to be adapted for each customer and for each country: XP could play its role here too.

The main obstacles against scaling up of XP are lack of documentation (therefore the exponential increase of necessary communication between developers), lack of stable interfaces and stable requirements. Consequently, scaling up of XP will probably be indispensable in order to adopt methodical practices from other methodologies.

In industry, a big reengineering wave started in the 80's, where companies tried to be more efficient regarding the competitor. From the discipline of systems engineering, we are acquainted with three approaches suited to manage a reorganization project. In the Total Systems Approach [Picot+79], the desired properties of a new system are defined first and then the whole system is introduced into the new organization like a big-bang invention. In the Incremental Systems Approach also known as "Muddling Through", a set of small changes incrementally leads to local optimization. Through small changes of the company structure and organization and its supporting software system, a series of small localized improvements lead to a sub-optimal organization form.

As both approaches have several drawbacks, discussed below, system engineering provides a third approach called "Hierarchical structuring" of system development or also "Mixed Scanning". This approach combines the advantages of both other approaches, usually leads to better reorganization projects and therefore to an overall optimization.

Some similarities exist between the extreme programming approach and the incremental systems developing approach, as well as similarities between the total system development and the classic OO methods approach. It is, therefore, an interesting issue to transfer the combination of these two approaches from systems engineering to the software engineering discipline. There are two basic advantages:

- the combination leads to a scale up of the extreme programming approach to larger projects by hierarchically structuring the teams, and

- it features a successful methodology named Genesis for organizing of hierarchical approach which can be transferred to the software engineering discipline.

Both of above discussed points have interesting aspects. Of course, scaling XP up to a larger project allows to apply the XP approach even if the system becomes more complex needing more people to be involved. Another advantage is that there is a proven methodology to get a hierarchical reorganization process organized; this can be easily adopted to software engineering discipline.

In Section 2 we shortly repeat the aspects of XP which are of interest in this context. We also discuss some of the advantages of XP over traditional OO methods to get a basis for the following discussion. In the third Section we introduce three different approaches to reorganize parts of the company and discuss their advantages and drawbacks. This will show the analogy to software process models and will point out some possible improvements for XP. In the fourth Section we discuss what XP can learn from company reorganization approaches and how a successful methodology (Genesis) can be used to incorporate XP.

## 2. XP - BRIEF INTRODUCTION AND DISCUSSION

Section 4 discusses adaptations of XP to a hierarchical variant. This is based on the principles of XP that we briefly introduce in this section. We suggest to skip this section, if the reader is familiar with the basics of XP. According to [Beck99], the light-weight methodology XP basically consists of:

- four values: communication between the programmers and between programmers and customers, simplicity of design, the early feedback through small changes in the releases and the courage of programmers to do whatever necessary,

- a number of software engineering principles discussed below,

- four basic activities: coding, testing, listening and designing, and

- a number of practices that help to structure the basic activities in order to achieve the basic principles.

Out of the four basic values a number of principles have been derived; this is again partially quoted from [Beck99]. Here are some fundamental ones:

- Rapid feedback: XP advocates very early a rapid, if possible, daily feedback that allows programmers to focus on the most important software features.

- Assumed simplicity: XP tries to focus on the simplest possible implementation that works. Therefore, it focuses only on today's problem and does not plan future extensions of software. In particular, it does not plan for re-use at all.11

- Incremental change: a big change will never work at the first try. XP advocates small changes to incrementally enhance the system with desired functionality. This idea is based on the concept of Refactoring, first introduced in [Opdyke+93].

- Embracing change: "The best strategy is the one that preserves the most options while actually solving your most pressing problem." ([Beck99], page 38)

- Quality work: quality is what finally matters. It emerges as dependent variable of the other three. The XP approach tries to ensure an excellent quality by focusing on basic principles that have proved useful and trying to keep the motivation of programmers up to its highest level.

For more on the software engineering principles, please refer to [Beck99].

There are four basic activities in the XP approach: coding, testing, listening and designing. As a light-weight methodology XP explicitly abandons any explicit activity of documenting and analyzing the system. Analysis remains a rather implicit, but continuous activity that happens in everyday communication with the customer. Documentation is explicitly discouraged. Therefore, it is not surprising that coding is one of the most important activities in the XP approach. The second important activity is the testing of written code. In order to ensure that the code written actually works, the XP approach advocates writing a high number of tests to check whether the code is correct or not. The test suit can be seen as a specification, some of them written by programmers themselves to ensure that programs they wrote actually work. In addition, customers or explicit testers also write functional tests to ensure the system meets the desired functionality. The third basic activity for XP programmers is listening. This means XP programmers listen to customer's needs and intentions as well as to those of other programmers by means of their everyday communication. Finally, even in the XP approach, it is sometimes necessary to step back from everyday coding and do some general design. This is of particular interest if changes to systems become more complicated.

---

[1]Disregarding re-use issues come simply from the observation that approaches that explicitly try to re-use software have not been too successful in this task. Furthermore, software seems to be best reusable if simple. So disregarding re-use may de facto lead to a higher re-use rate.

Even minor changes have impact on large parts of the system. A little bit of design is part of the daily business for all programmers in XP. Even though good design is difficult to evaluate, a number of hints exist. For example, the good design puts functionality that operates on data near to that data. It also tries to concentrate pieces of functionality within one place in system, not too distributed among numerous units. In order to establish all the goals XP advocates, the good design also gives us several practices that should be followed:

- *The planning game*: based on business priorities and technical estimates, the scope of the next release should be determined. Please note that releases depend slightly on one to a few of their incremental developments.

- *Small releases*: one release almost every day or a couple of days.

- *Simple design*: systems should be designed as simple as possible. This means in particular, that unused functionality that complexes the system further should be removed.

- *Testing*: unit tests are written by programmers, functional tests are written by customers to demonstrate that a feature actually works.

- *Refactoring:* the techniques of refactoring deal with structural changes of a system without affecting its behavior. They simplify the system, remove unused code, add flexibility or improve the communication between the parts. This can be done, e.g. by splitting or joining classes or moving methods or attributes from one class to another [Opdyke+93].

- *Pair-programming:* two programmers sit at one machine. What one programmer writes is immediately reviewed by the other. They continuously communicate with each other to ensure directly the high quality of their code.

- *Collective ownership:* every programmer may change each part of the program at any time. There is no single owner of the code.

- *Continuous integration:* the idea is to integrate the new code and build the system many times a day whenever a task is completed. Of course, each time the set of all tests is checked against the new version of the system.

- *On-site customer:* in order to ensure continuous communication not only among programmers, but also with at least one customer, it is important to have one customer in the team all the time. He can immediately answer all the questions that programmers have.

- *Coding standards.*

For a more complete list and a detailed list of these practices, please refer to [Beck99].

Today, the XP approach is mainly used for small projects (1-10 team members). It has been successful within small-scale projects, but had not really been applied to larger projects. Due to its basic values and principles, we may assume that XP in its current form does not easily scale up to larger projects. However, if we compare the XP approach to classically oriented methods, like the already mentioned Catalysis, the Unified Process or the methodology framework provided by OPEN, we discover several advantages of XP. Compared to classical analysis design implementation or to more modern Inception-Elaboration-Construction-Transition-life-cycle, XP is more flexible and includes more explicitly the needs and intentions of all project participants. As in XP the stages of analysis, design and implementation are no longer separated, the analyst and the implementer are basically united in a single person. This, on one hand, allows us to forget about a thorough analysis documentation. On the other hand, the implementer is directly in touch with the customer, and, therefore, knows his intentions thoroughly when mediated by the analyst. Such considerable increase of motivation is paired to the appropriate decrease of misunderstandings on the communication path from the customer to the implementers.

Altogether, XP is a promising approach with a number of strengths and few weak spots that can be improved. This paper presents an approach to tackle the specific problem of scaling XP up to larger projects. The characteristics of XP's advantages and disadvantages being similar to different ways of reorganizing companies, like those used by the automobile industry in the 80's and 90's, we will show in the next chapter the improvements made when reorganizing companies. This gives us more hints to improve XP.

## 3. THREE APPROACHES TO COMPANY REORGANIZATION

Many different, more or less successful approaches were developed to reorganize companies in the past. A systemization of these approaches suited for a reorganization of companies divides them into three different types [Veitinger95]:

- *The Holistic Approach*, also called Total Systems Approach

- *The Incremental Approach,* also called Muddling Through

- *The Hierarchical Approach*, also called Mixed Scanning or Piecemeal Scanning

We shortly introduce each of these in its own right, and then compare them to each other.

### 3.1    The Holistic Approach

The holistic approach tries to manage the whole complexity of a planning process in one comprehensive step, and does not divide the problem into any smaller parts. With this approach, it is necessary to understand the whole problem in detail before working out a solution. This needs a broad information basis and a detailed as-is-analysis. To initialize a reorganization process, one also has to tag one's targets exactly. Only then, various alternatives can be generated and the best-fitting alternative chosen.

For a detailed as-is-analysis of an existing system in a complex environment, it is sometimes useful to structure it into subsystems. Each subsystem with the functionality (described through visible input and output) will then be analyzed separately. After the as-is-analysis, we can again regard those subsystems as black-box description of a system. This is basically the same approach as if a new system is being built by starting with a rough planning on a high level, treating all subsystems of lower level as black-boxes. The holistic approach is typically used in operation research, statistical decision theory and systems analysis. When trying to reorganize more complex systems, certain difficulties occur:

- The capacity of solution solvers is limited for a detailed analysis of all possible solutions and their consequences.

- The derived description, i.e. information-flow does not fulfill the requirements.

- The process is time-consuming and costly (partly because of the detailed analysis phase).

- Provided the system is open to future change, all economic influences must be precisely estimated.

A significant disadvantage remains the enormous need of high-level competence and detailed holistic knowledge to work out a solution. Numerous reorganization projects of complex systems failed only because the solution was not reached in given time.

## 3.2    The Incremental Approach

The incremental approach develops the system step by step. The primary target is not a perfect solution for a problem but a continuous progress. Consequently, the real targets of reorganization exist only on a very high level. It therefore must be checked periodically whether the activities are still effective and meet the targets. A typical example of an incremental approach is the continuous improvement process (KAIZEN, originated in Japan). It is based on the assumption that all steering and control mechanisms can be changed independently. Six principles characterize the incremental approach:

- Comparison and Analysis are done step-by-step: the reorganization process starts with the actual system and tries to improve it in small steps. The problem solver focuses on the well known elements.

- The number of alternative solutions is restricted: possible solutions are restricted since the modifiable elements stand only for a small part of the whole system.

- Restricting the regarded consequences: as we can modify only a small part of the system, the consequences of a problem solution may regard only these few elements; otherwise, the changes are not acquainted with context.

- Simplification: make a complex problem easy by dividing it into several smaller ones and solve them first (but independently) before going further to a final solution.

- Repeating analysis and evaluation: a certain situation may always be improved by finding a better solution than the existing one.

- Healing negative symptoms: fast removal of weak points of a big complex system, without trying to develop overall solutions. This approach is also called inductive.

The incremental approach is easy for use and of great help when quick improvements for small elements are needed. This is achieved by only involving the employees who best know their processes. The disadvantages of this approach are the lack of innovative solutions and the lack of one common target.

3.3    The Hierarchical Approach

Both the holistic and the incremental approach are combined here. Starting from an overall view of a reorganization project, the whole system is divided into modular, very detailed elements. When we work on such elements, we are always able to give feedback on a higher conceptual level in order to check targets and singular planning. Several principles should be fulfilled to make this approach successful:

- The problem should be clearly defined by an input / output relationship and furthermore divided into elements. Possible criteria for the division are processes, products or functions.

- Clearly defined targets for each element are needed.

- The target is approached stepwise through improvements on each particular step.

- A special form of project-management is necessary to handle the communication and coordination of the modular elements appropriately.

Experience shows, especially the coordination task is a great success factor when it comes to practical solutions that fulfil the targets adequately.

The analysis of each of the three approaches shows various ways to undertake a reorganization process. All have been used in practice. The holistic approach is normally implemented when changing bigger organization structures, if personal interests are involved. Then a solution in such case cannot be reached involving the employees, since it takes incredible planning time and personal objections are then bad driving forces. With personal objections involved there is a high risk of not reaching the goals at all. The incremental approach, on the other hand, is chosen in case of highly complex structures without an overall detailed reorganization concept. There are only high level targets, which do not allow explicit suggestions on the operational level. Last but not least, the hierarchical approach has become a common choice for reorganizing companies. It allows to combine the power and knowledge of employees on the detailed level while solving problems with innovative concepts. Targets are well-defined and a special coordination function is established in the project-management. To satisfy all principles of an hierarchical approach, one basic method named Genesis has been brought to light, with great success [Wildemann 97]. This method helps by means of coordination to reorganize all elements in a system with defined targets and is described below.

3.4     Genesis: Managing the Change

Genesis is a management concept that consists of a combination of proper tools and techniques to improve the objectives of a business process and that also includes some management elements to guaranty the success. It is used to simplify work, taking advantage of a very efficient cross-functional team-concept without many levels of authorities and to operationalize a strategy for managing a change. Genesis is a way to reach and implement quickly solutions in a defined process with inputs and outputs. The Genesis approach:

- supports readiness to change things,

- encourages every subordinate to check one's personal field of action or obstacles and to use this evaluation,

- causes big impact implementing in a lot of singular steps,

- solves problems which can be handled by the subordinates themselves,

- helps to realize partial solutions with small efficiency,

- activates creativity, the idea pool and the performance of all subordinates, and

- encourages employee's responsibility for the own task area.

The Genesis hierarchical approach can be organized as a series of workshops with a clearly defined program for the reorganization object. It starts with a preparation phase, where the project organization, the high level targets and the objectives of this reorganization are defined. Usually, all team-members working in project are taught the same problem-solving methods in order to have one common language. After setting up the rough structure and the project team, a first planning workshop starts. Participants of this planning workshop usually come from the management. The management defines the basics of the new, innovative concept and then divides the system up into its elements. To add to it, all targets for the elements and their process-input and output are defined in the first workshop. For each element, a new workshop is defined. To coordinate all these workshops, there is a core team in the project organization with special tasks, meeting regularly. In these meetings, problems are discussed, e.g. whether a solution for one element effects another. Such problems are first checked by members of the project-organization group. Ideally, discussions concerning possible effects on other elements should not arise at all, otherwise the whole system was divided into wrong elements.

Numerous case studies show that the hierarchical approach has proved to be the most successful for reorganizing companies, whenever very detailed information is necessary with respect to the processes, whereas at the same time though, implementation of new innovative concepts is desired. At this point, the holistic approach did not yield acceptable solutions, as an entire, undivided system was too complex to handle. The incremental approach also did not achieve the targets set, because each solution for an element can be perfect in its own right but not always perfect for the whole system. The key method to reorganize companies with the hierarchical approach was the Genesis method. The success of this method was shown in more than 500 workshops with an average saving of about 250,000.- $ [Wildemann97]. The success factors of the Genesis method are:

- involvement of all employees, also top management and management,

- defined communication steps, e.g. like a top-management presentation after four days to put pressure on finding solutions,

- defined communication concept and coordination of multiple teams,

- supporting a clearly defined target through all workshops and team activities.

The description of the different approaches show that the hierarchical approach combines the advantages of the holistic and incremental approach thus eliminating their disadvantages. The major success factor of the hierarchical approach is a management process, that is described by Genesis. In the following, we will try to transfer this kind of improvement to a new way for XP with a hierarchical approach.

## 4. COMBINING XP AND THE HIERARCHICAL APPROACH

In the previous sections, we witnessed a characterization of the XP approach as a light-weight software methodology, and the three variants for company reorganization. Today it is for all companies imperative for success in a competitive market to supply their business with extensive software support. A company reorganization always goes together with the adaptation of existing and the introduction of new software. Therefore, it is a natural consequence to combine suitable approaches that come from technical and management disciplines, particularly, if both approaches possess a structure with similar characteristics. Therefore, the first step towards such a combination can be a comparison of the discussed approaches. Despite existing differences, a comparison also reveals some similarities between the following approaches in software development resp. company reorganization:

| Software Development | Company Reorganization |
|---|---|
| Classic Software Engineering (Waterfall and Heavy-Weight OO) | Holistic Approach |
| Extreme Programming | Incremental Approach |
| Hierarchical Extreme Programming | Hierarchical Approach |

The holistic approach has several characteristics in common with the classical software engineering approaches, starting with the Waterfall model, but also newer object-oriented approaches, like the Unified Process [Jacobson+99]. They e.g. share a centralized approach providing a small coordination team with great power, but lack adequate customer/employee participation. Their major advantages are:

- a common, clearly to be identified target, and

- a structured process that leads the team through the software development phases.

The major disadvantages:

- lack of involvement of employees in charge of the holistic approach, whereas in the object–oriented approaches, the respective customers play a minor role. We miss in particular the early feedback of the end-users.

- the missing customer/employee participation leads to reduced acceptance of the solution. It is far easier to convince the people of new ideas, when they actively helped to shape them.

- The incremental approach is comparable to XP. Both are rather decentralized and both focus minor on local improvements of existing structures/systems. Such improvements can be released early and get a fast feedback. Their major advantages are:

- high involvement of employees / customers, and as a result,

- high acceptance of the solution.

- XP and the incremental approach do have also some disadvantages in common:

- applying this approach to several local problems does usually not lead to a shared improvement with multiple teams. Instead, local improvements may contradict each other,

- the approach is unstructured and cannot therefore be used for working out an overall concept by a complex problem where an involvement of several persons is necessary.

One interesting question is whether the hierarchical approach, as a combination of the holistic and incremental approaches, can be carried over to the software development discipline. Currently, newer object-oriented approaches, like the Unified Process incorporate some of the elements of the hierarchical approach. However, these approaches are too heavy-weight to fully support the advantages of the hierarchical approach. Therefore, we decided to start from the extreme programming approach and build a hierarchical structure upon it. The advantages of this approach have partly been discussed before: While retaining a light-weight methodology, it becomes feasible to structure larger projects into a bunch of smaller XP projects that still have a common target to achieve. The approach taken also fits nicely into the Genesis program, which basically consists of two important elements:

- on the top-level we set up a goal-oriented project management that organizes the problem as a high-level structure by working out a rough concept,

- each of the now localized problem parts is solved in an extreme programming approach by defining an own XP team.

Figure 1 demonstrates this advantage of the hierarchical approach compared to the other two approaches. Each circle is a tea member, tight connections of circles form a team.

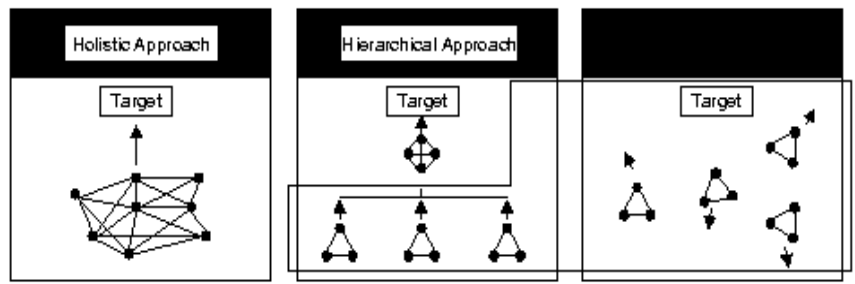

**Figure 1:** Comparing the three approaches of company reorganization

The XP teams function primarily on an independent basis; nevertheless, they are coupled in a slack manner by a top-level management team, called "steering committee", that keeps track on the overall goal and measures local improvements. Even though the XP teams dealing with local problems are to be formed as principally independent units, some cross-dependencies usually arise. It is important to keep these cross-dependencies as lean as possible. However, dealing with the complexity of today's information systems, this is a difficult issue. Today software engineering techniques still do not sufficiently provide mechanisms to define crisp and small interfaces between software parts (and their development projects). If the actual initial problem structure and its software solution turns out to be inadequate, an appropriate refactoring of this structure is encouraged. This goes along with dynamic restructuring of the XP teams, which is useful anyway to dynamically react on varying workloads. Refactoring at this level is surely more expensive, as it crosses team boundaries. Therefore it is imperative to have the top-level project management at hand to handle these kinds of restructurings.

By organizing the software development process in a hierarchical manner we always focus on one common target and use a structured process to reach this goal. The involvement of the employees leads to a high solution acceptance starting from middle up to highly complex project settings. Ideally, XP project teams are defined similarly to company departments. The XP project team structure, therefore, resembles the company structure. Thus, each XP team recruits its customer(s) from its associated department(s).

A certain part of the software infrastructure of a company is not localized in one (or a few) departments, but its usage is spread over a number of departments. Of course its development or enhancement is still a matter of one XP project, but if several of these software subsystems are to be restructured or enhanced, then the setup of the XP project structures in analogy to departments structures is not appropriate anymore. This can partly be solved by identifying pilot departments with pilot customers that are able to cover the needs and desires of other departments' users as well.

These considerations show that the hierarchical extreme programming approach needs a good, yet lean project organization. This is supported by the Genesis program and can be carried over to software project structures. Five major principles can be identified that characterize the hierarchical approach (cf. Figure 2).

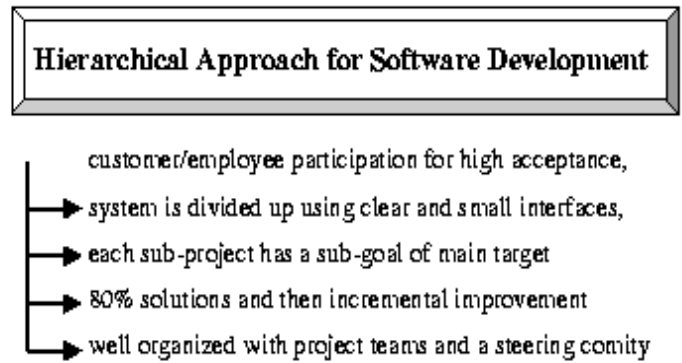

**Figure 2:** The five major principles of the hierarchical approach

These principles are a combination and adaptation of corresponding XP principles and principles from the hierarchical reorganization approach.

1. Customer participation: the solution is worked out with the customer/employee to reach a high acceptance. This is in particular important for the customers to accept the resulting new software system / company structure.

2. The whole system is divided up into subsystems with a lean and crisp interface. The inputs and outputs, namely the data structures and the information flow between the subsystems need to be clearly defined. Subsystems are implemented respectively evolved through XP teams.

3. Each XP team has as target its associated subsystem, thus contributing to the main target, namely the development of the whole system.

4. The worked out software solutions will be improved like an incremental process to be successful very quickly. In a number of releases the team explores and extends the desired system functionality.

5. The hierarchical approach needs to be well organized with a project team and a steering committee followed by a Genesis program. The steering committee is an ideal place to develop and maintain the common system metaphor. This cannot be decentralized in several XP teams, but needs a central coordination.

Most of the additional principles and practices of XP, that have been introduced in Section 2, carry over to the hierarchical approach without major changes. However, some of these principles need slight enhancement. The test suite becomes even more important when the XP teams have interfaces to each other. Tests need to be automatically conducted that check the correctness of the cross project functionality and therefore the correctness of the interfaces.

**Case Study: Bidding Software**

As a first case study (in non-hierarchical XP) in a management circle "Electronic Sourcing" held at the Technical University of Munich we decided to develop a software able to support different parts of the purchasing process. This management circle was joined by purchase managers of e.g. BMW, Dresdner Bank, DaimlerChrysler, Philipps etc. To find out what features should be developed, we started with a planning workshop with mainly customer and consultant participation. In order to systemize purchasing processes we started with identifying existing needs and continued with controlling the purchasing activities. Different concepts like online broker, purchasing card, and online shopping malls where evaluated and online bidding were identified as the most valuable techniques to reduce costs. This workshop was dominated by the participants of the management circle and members of the business and economics institute. In a second workshop, where also information technology people were involved, we defined the core features of the B2B real-time online bidding software. In a first (non-hierarchical) XP project, the online bidding software was developed. All elements of the software were developed using an incremental approach with short feedback loops to the customer (intensive meetings 2-3 times a week, weekly releases). In this way we were able to develop a bidding software in a very short time with a high acceptance for the customers, namely the consultants using the software to actual B2B online bidding. The software started productive use after only 8 weeks of development, but is still evolving through adding flexibility and new functionality. This evolvement is necessary to ensure meeting the needs of consultants and their customers that cannot be foreseen in this highly innovative field of e-commerce. With this software we realized an impressive number of auctions with a volume of millions of Euro in two months with an average saving of approximately 20%.

In a second and larger effort, extension subprojects have been defined that also use the XP process for their development. An overall steering committee is ensuring that all sub-projects stay on track. A second case study dealing with hierarchical XP is currently under investigation. The results will surely take time, as a hierarchical approach to XP cannot be implemented in field as fast as XP can.

## 5. CONCLUSION

In the context of the FORSOFT project and its industrial partners, we had a close look at the basic question of software engineering. FORSOFT focuses both theoretical and practical issues of software development as well as management issues, in particular, on their combination. This paper contains a result that focuses on the particular question how to optimize and reorganize companies that make heavy use of software products.

If a company is reorganized, this usually means restructuring its software products, its databases and network infrastructure, because a company reorganization usually concentrates on optimization of its business cases. In this paper, we have introduced and discussed different software development processes as well as different company reorganization approaches. In particular, we have discussed similarities and differences between these approaches and concluded that an extension of the extreme programming approach by elements of the hierarchical reorganization process leads to a considerable scale-up of the XP approach. Furthermore, XP approach extended this way can nicely be integrated with the hierarchical reorganization process allowing the use of both at the same time.

Another interesting issue for further discussion is the fact that the hierarchical approach cannot oddly be carried over to software development discipline. Usually both disciplines – reorganizing a company and restructuring the software – go hand in hand. Therefore, it turns out to be extremely useful to combine the hierarchical management approach with above software development approach in a hierarchical software and company restructuring setup. Thanks to integrated project structure, we can usually expect far better results than if we use the centralized approach, which does not involve employees/clients enough, or a decentralized or unstructured approach that does not focus sufficiently the global target.

## Acknowledgements

This work was partially supported by Siemens AG Munich, the Bayerische Forschungsstiftung under the FORSOFT II research consortium, and the Bayerisches Staatsministerium für Wissenschaft, Forschung und Kunst under the Habilitation-Förderpreisprogram.